\documentclass[twocolumn,10pt,prl,nofootinbib,preprintnumbers]{revtex4-1}
\usepackage[utf8]{inputenc}

\def\mysection#1{{\bf #1.} }
\def\mysections#1{{\bf #1.} }
\usepackage{hyperref}
\usepackage{cleveref}
\usepackage[dvips]{graphicx}
\usepackage{feynmp}
\usepackage{siunitx}
\usepackage{epsfig,amsmath,amssymb,verbatim,mathrsfs,array,layout,textcomp,amssymb,latexsym,slashed,graphicx,booktabs,color,mathtools,tikz}

\newcommand{\beq}{\begin{eqnarray}}
\newcommand{\eeq}{\end{eqnarray}}

\def\beqa{\begin{eqnarray}}
\def\eeqa{\end{eqnarray}}

\newcommand{\bv}{\left(\begin{array}{c}}
\newcommand{\ev}{\end{array}\right)}
\newcommand{\bmtwo}{\left(\begin{array}{cc}}
\newcommand{\bmthree}{\left(\begin{array}{ccc}}
\newcommand{\emn}{\end{array}\right)}
\newcommand{\bmtwoc}{\left\{\begin{array}{cc}}
\newcommand{\bmthreec}{\left\{\begin{array}{ccc}}
\newcommand{\emnc}{\end{array}\right\}}
\newcommand{\ba}{\begin{array}}
\newcommand{\ea}{\end{array}}

\def\lsim{\mathrel{\rlap{\lower4pt\hbox{\hskip1pt$\sim$}}
     \raise1pt\hbox{$<$}}}         
\def\gsim{\mathrel{\rlap{\lower4pt\hbox{\hskip1pt$\sim$}}
     \raise1pt\hbox{$>$}}}         

\begin{document}
\font\mini=cmr10 at 0.8pt

\title{
Detecting Dark Matter with Superconducting Nanowires
}

\author{Yonit Hochberg${}^{1}$}\email{yonit.hochberg@mail.huji.ac.il}
\author{Ilya Charaev${}^{2}$}\email{charaev@mit.edu}
\author{Sae-Woo Nam${}^3$}\email{nams@boulder.nist.gov}
\author{Varun Verma${}^3$}\email{varun.verma@boulder.nist.gov}
\author{Marco Colangelo${}^{2}$}\email{colang@mit.edu }
\author{Karl K. Berggren${}^{2}$}\email{berggren@mit.edu}
\affiliation{${}^1$Racah Institute of Physics, Hebrew University of Jerusalem, Jerusalem 91904, Israel}
\affiliation{${}^2$Massachusetts Institute of Technology, Department of Electrical Engineering and Computer
Science, Cambridge, MA, USA}
\affiliation{${}^3$National Institute of Standards and Technology, Boulder, CO, USA}

\begin{abstract}

We propose the use of superconducting nanowires as both target and
sensor for direct detection of sub-GeV dark matter. With excellent sensitivity to
small energy deposits on electrons, and demonstrated low dark counts,
such devices could be used to probe electron recoils from dark matter
scattering and absorption processes. 
We demonstrate the feasibility of
this idea using measurements of an existing fabricated tungsten-silicide nanowire
prototype with 0.8-eV energy threshold and 4.3~nanograms with 10 thousand seconds of exposure, which showed no dark counts. The results from this device already place
meaningful bounds on dark matter-electron interactions, including the strongest
terrestrial bounds on sub-eV dark photon absorption to date.
Future expected
fabrication on larger scales and with lower thresholds should enable probing new territory in
the direct detection landscape, establishing the complementarity of
this approach to other existing proposals.
\end{abstract}

\maketitle

\section{Introduction}

Dark matter (DM) is one of the most important unsolved mysteries of the Universe. Focus on the
Weakly Interacting Massive Particle (WIMP) paradigm has guided
experimental searches for decades. Traditional methods searching for
such weak-scale DM in the laboratory via nuclear recoils have made
tremendous progress is probing DM with mass above the GeV-scale, but
typically make poor targets for detection of sub-GeV DM that goes
beyond the WIMP paradigm. As the WIMP parameter space continues to be
covered without discovery of DM, new ideas to search for lighter DM
are of the essence.

Indeed, recent years have seen a surge of such new ideas
emerge. These include the use of atomic
excitations~\cite{Essig:2011nj}, electron recoils in
semiconductors~\cite{Essig:2011nj,Essig:2012yx,Graham:2012su,Kurinsky:2019pgb},
2-dimensional targets such as graphene~\cite{Hochberg:2016ntt} and
carbon nanotubes~\cite{Cavoto:2017otc}, color
centers~\cite{Budnik:2017sbu} and
scyntillators~\cite{Derenzo:2016fse}, which can be sensitive to
MeV-scale DM masses. Sub-MeV DM can further be probed by
superconductors~\cite{Hochberg:2015pha,Hochberg:2015fth,Hochberg:2016ajh},
Dirac materials~\cite{Hochberg:2017wce}, superfluid
helium~\cite{Schutz:2016tid,Knapen:2016cue} and polar
crystals~\cite{Knapen:2017ekk,Griffin:2018bjn}.  The proposed
experimental designs for each distinct target material differ from one
another, with a variety of sensor technology employed across designs,
including the use of CCDs, TESs, MKIDs and G-FETs coupled to a target.

Quantum information science has been breaking new ground in sensor
technology, with superconducting nanowires a now established and
burgeoning field~\cite{Natarajan2012, Hadfield2009, Najafi2016}.  Some
of these nanowires have sub-eV energy sensitivity, which allows them
to be used as single-excitation detectors.  The recent emphasis on
development of such low-threshold, ultra-fast and low-noise
single-photon detectors for photonic quantum information
applications~\cite{Takes2007,Calkins2013} promises a radical
improvement in the search for DM.  The advent of superconducting
nanowire detectors, which currently have fewer than 10 dark counts per
day~\cite{Wollman2017} and have demonstrated sensitivity from the
mid-infrared~\cite{Marsili2012} to the ultraviolet wavelength
band~\cite{Wollman2017}, provides an opportunity to search for rare
low-energy deposits of DM via scattering or absorption processes.

Here we propose and perform initial experiments using this technology as both the target for DM
interactions with electrons and the sensor with which to detect
these interactions. Depending on the energy thresholds reached in
these devices---nanowires with sub-eV thresholds have already
experimentally realized~\cite{Marsili2012}---sensitivity to low-mass DM
can be achieved. Energy deposits of order a few eV and above can
further allow for directional detection of DM via a stacked geometry,
which would serve as a powerful discriminate between signal and
background.

In this letter, we begin by describing the basic detection process in
these devices. We then describe an existing prototype nanowire and
report on how it can be used to extract new bounds on DM interactions with
electrons both in scattering and absorption processes. 
Our projections for future reach of superconducting nanowires into the DM
parameter space follow. We conclude with discussion of impact, remaining
issues and possible future work.

\section{Concept}

\begin{figure}[t!]
\includegraphics[width=.48\textwidth]{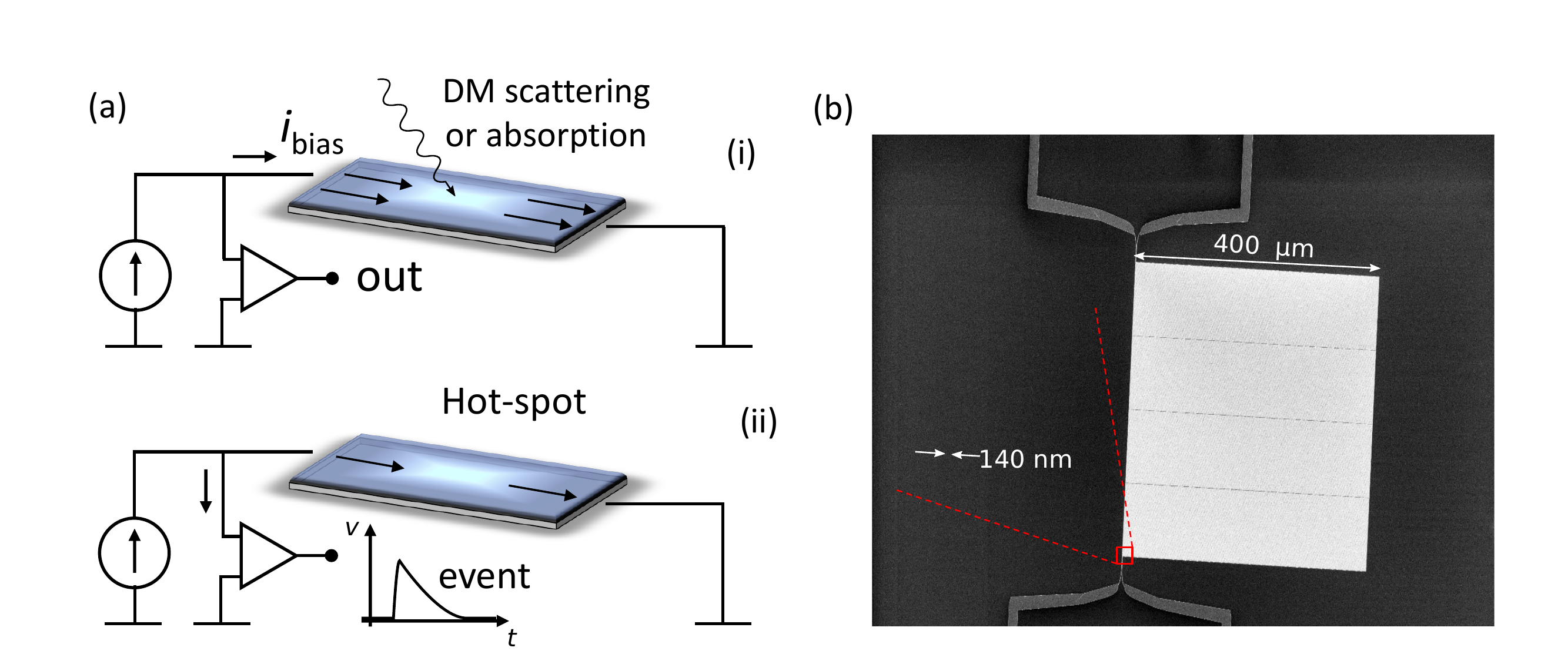} 
\caption{\label{fig:snspd} (a)~Schematic depiction of the operating
  principle of SNSPDs: (i)~The detector is biased at a current close
  to the critical value. (ii) When the energy is absorbed by the
  nanowire, the electrons depart from equilibrium and diffuse out of
  the formed hot-spot. A resistive region formed across the nanowire
  then leads to a measurable voltage pulse in the readout. (b)~The SEM
  image of the prototype WSi device after fabrication. The active area
  is 400 by 400 $\mu$m$^2$. Nanowires are consistently connected to
  two contact pads.  }
\end{figure}

Superconducting nanowires are a rapidly developing technology with
applications ranging from space communications~\cite{Grein2014,
  Grein2014b}, to LIDAR~\cite{McCarthy2013, Chen2017}, to
quantum-information science \cite{Natarajan2009}. With sub-eV energy
sensitivity~\cite{Marsili2012}, $\sim 10^{-4}$ counts per second~(cps) dark count
rates~\cite{Wollman2017}, and spatial discrimination
ability~\cite{Zhao2017}, superconducting-nanowire single-photon
detectors (SNSPDs) provide an excellent candidate for detecting
DM. SNSPDs are fabricated by using superconducting films a few-nm
thick on a variety of substrates, with widths between 30~nm and 200~nm
using electron-beam lithography and reactive ion etching. SNSPDs are
typically fabricated into a planar meander structures covering tens to
hundreds of square micrometers~\cite{Lv2017}. The device operating
principle is straightforward: when cooled below the superconducting
transition temperature and biased with a sufficiently high current,
the energy deposited by an incident particle can cause the transition of
a portion of the nanowire into the normal (resistive,
non-superconducting) state. This appearance of a resistive region in
the current-biased nanowire results in voltage pulses
with typical amplitudes of $\sim\SI{1}{\milli\volt}$ (depending on the
amplifier's input impedance), and durations of a few to tens of
nanoseconds. A schematic depiction of the device operation is shown in
Fig.~\ref{fig:snspd}(a). These detectors have demonstrated dark count
rates as low as 1$\times 10^{-4}$ cps~\cite{Wollman2017}, making them particularly interesting for
sensing rare events.

We therefore propose the use of SNSPDs for direct detection of
DM. They can be used as both the target material with which the DM
interacts, as well as the sensitive sensor measuring this
interaction. Large target mass can be achieved via large arrays
combined with multiplexing~\cite{Dauler2009}, without disturbing the excellent energy
threshold of these devices nor their low-noise character.

A useful rule-of-thumb regarding the connection between the energy
threshold of the device versus the DM mass that it can probe is as
follows. In a DM scattering process off a target, the maximal energy
deposited is the entire kinetic energy the particle is carrying $\sim
m_{\rm DM} v_{\rm DM}^2$, with $m_{\rm DM}$ and $v_{\rm DM}$ the DM
mass and velocity, respectively. Since the DM velocity around us
is of order $10^{-3}$ in natural units (where $c=\hbar=1$), a given system sensitive to
energy deposits of $E_{\rm D}$ or larger can probe DM masses $10^6$ larger
than $E_{\rm D}$ via the scattering process, $E_{\rm D}^{\rm scat}\sim
10^{-6} m_{\rm DM}$. If instead the DM particle is absorbed by the
target, it deposits its entire mass-energy, meaning that the same
target system is sensitive to $E_{\rm D}^{\rm abs}\sim m_{\rm DM}$ via
absorption processes.

For DM scattering with electrons in the SNSPDs, devices with eV-scale
thresholds can thus probe DM mass of MeV and above. In this mass
range, several proposed other targets exist in the literature (see,
{\it e.g.}, Ref.~\cite{Battaglieri:2017aum} for a recent community
report).  The reach of the SNSPDs can be comparable to or better than
these other targets, depending on exposure size and duration, and is complementary
to other approaches. The SNSPDs, however, offer the advantage of
possible directionality of the signal: with energy deposits of a few
eV and above, the electrons are likely to be ejected from the
material, and could then hit multiple layers of SNSPD arrays.  If it
is found that the ejected electron from the superconductor tracks the
direction of the incoming DM particle~\cite{SCdir}, then reproducing
the direction of the outgoing electron via the stacked geometry and the SNSPD's spacial discrimination power would
inform us about the directionality of the signal. This could also
help discriminate signal from background. Similar use of
directionality from a stacked configuration has been suggested for use
of in graphene targets~\cite{Hochberg:2016ntt}.

\begin{figure}[t]
\includegraphics[width=.48\textwidth]{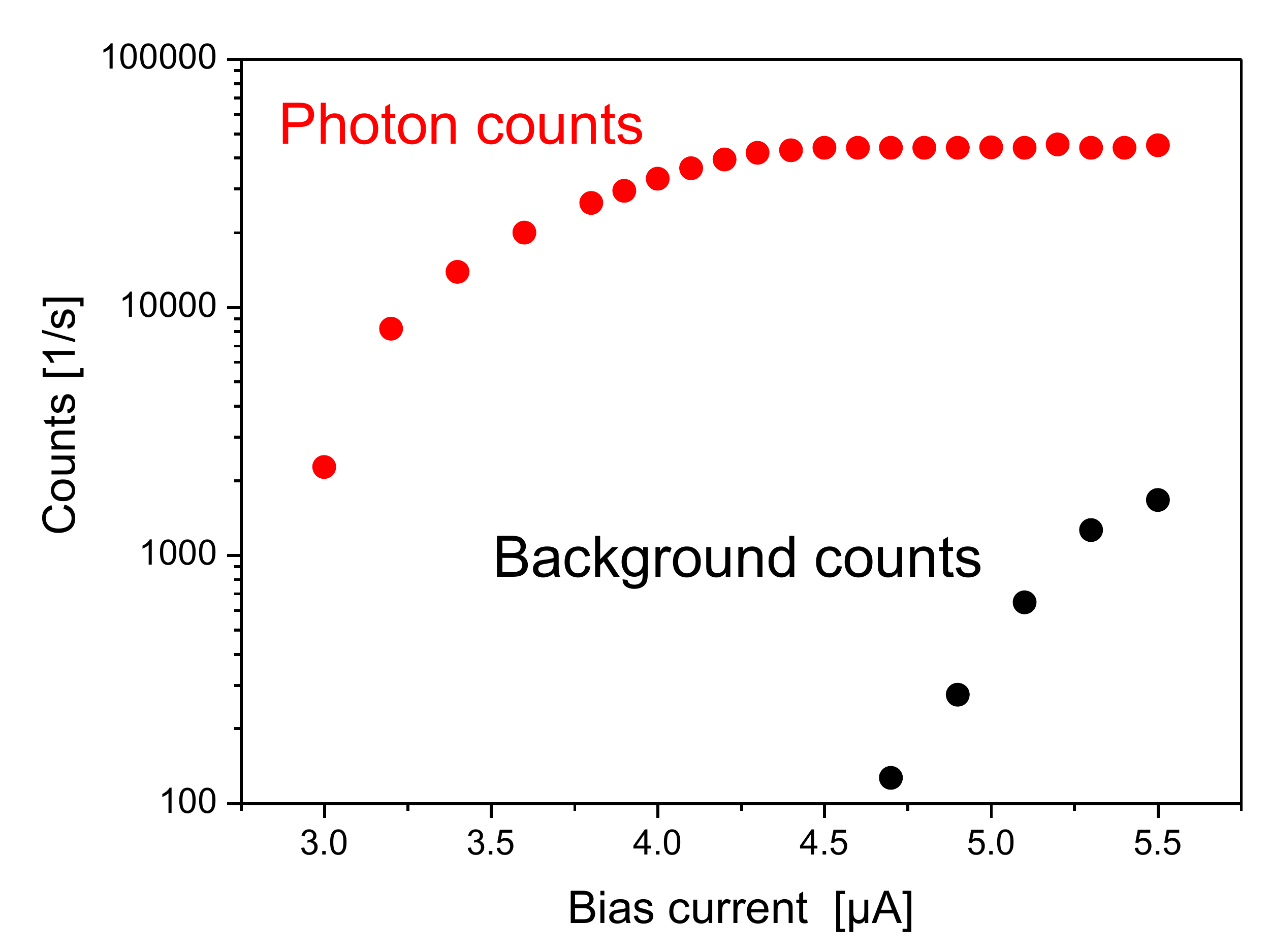}
\caption{\label{fig:pcr} The photon (red) and background (black)
  counts as a function of the absolute bias current, exhibited by the prototype WSi device tested in a fiber-coupled package at \SI{300}{\milli\kelvin}. }
\end{figure}

\begin{figure*}[t!]
\includegraphics[width=0.48\textwidth]{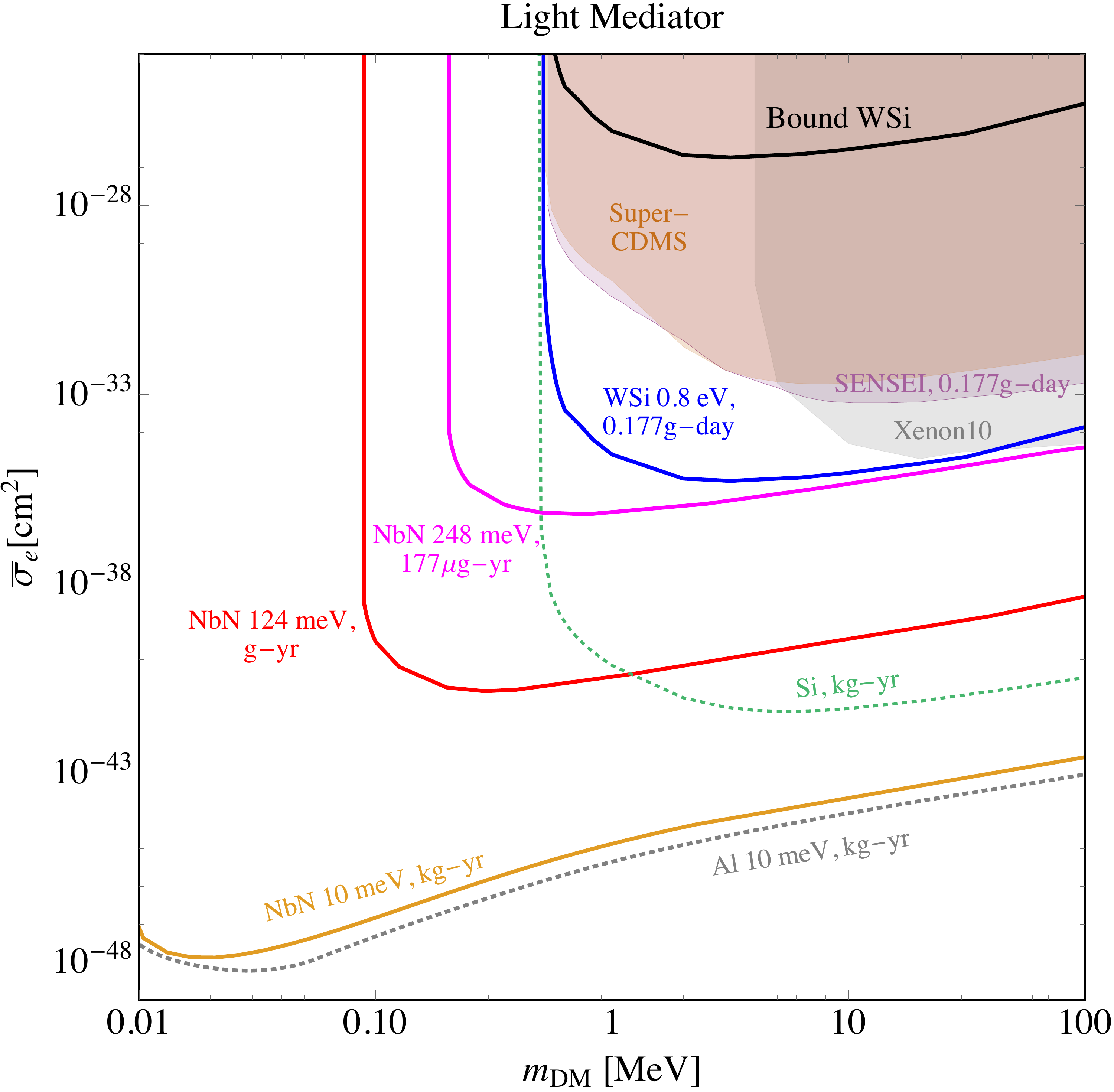} 
\includegraphics[width=0.48\textwidth]{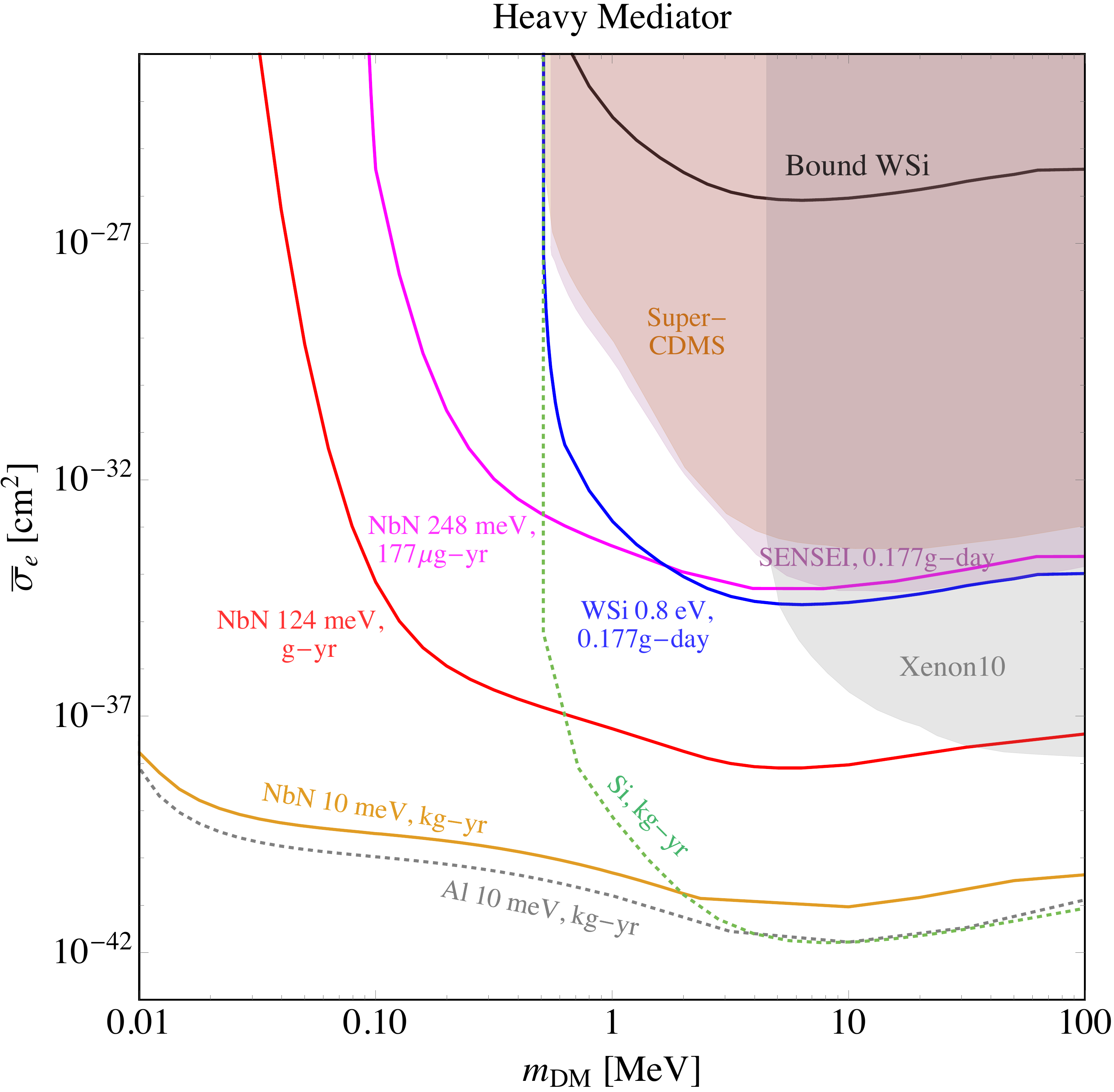} 
\caption{\label{fig:scat} Expected reach for DM-electron scattering
  via a light ({\it left panel}) and heavy ({\it right panel})
  mediator as a function of DM mass.  The solid black curve labeled `Bound WSi' indicates the
  new bound placed by our prototype device with 4.3~ng exposed for 10,000 seconds. Other solid curves
  indicate our 95\% C.L. projected reach for either NbN or WSi
  targets, with various exposures and thresholds. Also shown are the existing
  constraints from Xenon10~\cite{Essig:2012yx} (shaded gray),
  SuperCDMS~\cite{Agnese:2018col} (shaded red) and
  SENSEI~\cite{Abramoff:2019dfb} (shaded purple), as well as the projected
  reach for a kg-yr exposure of a silicon target~\cite{Essig:2015cda} (dotted green)
  and superconducting bulk aluminum with a 10~meV
  threshold~\cite{Hochberg:2015pha,Hochberg:2015fth} (dotted gray).  For clarity, 177~$\mu$g
  corresponds to a 10~by~10~cm$^2$  area of NbN at 4~nm
  thickness and a 50\% fill factor, and 248~(124)~meV threshold
  corresponds to a 5~(10)~$\mu$m wavelength.  }
\end{figure*}

As the threshold of the device is lowered to sub-eV energies, lower DM
masses can be probed, with ${\cal O}({\rm meV})$ energy deposits above the superconducting gap corresponding to ${\cal O}({\rm keV})$ DM
masses. Indeed, nanowires that exhibit sensitivity to \SI{5}{\um}
wavelength photons, corresponding to an energy threshold of
$\sim\SI{250}{\milli\electronvolt}$, have been
demonstrated~\cite{Marsili2012}, and it is likely that further
technology developments could push the energy sensitivity to
\SI{10}{\um} or even beyond. As we will show, the reach of the SNSPDs
into the sub-MeV DM mass range is substantial, and can provide
excellent results even with very small target masses, which can be
constructed on relatively short time scales.

Additionally, as we will show, absorption of DM in the sub-eV and
above mass range is similarly possible via SNSPDs, providing an
important complementary probe to {\it e.g.} existing stellar
constraints.

\section{Existing Prototype Device}

Having presented the basic concept of detection via SNSPDs, we now
describe an existing prototype device and how measurements of its
performance already place bounds on DM scattering and absorption.

Fig.~\ref{fig:snspd}(b) is a scanning electron micrograph (SEM) of the prototype
tungsten silicide (WSi) device after fabrication. The active device
area was 400 by 400~$\mu$m$^{2}$, and the nanowire was connected to
external circuitry via two contact pads. The width of the nanowires
was 140 nm with a pitch of 340 nm. The thickness of the WSi film was 7~nm, and the resulting mass is 4.3~ng. Further details of the device design and fabrication
are provided in the Appendix.

The switching current of the device $I_C$ was 5.5 $\mu$A
was measured at 300~mK by sweeping the current from a \SI{50}{\ohm}
impedance source. Fig.~\ref{fig:pcr} shows the dependence of the count
rate on the absolute bias current for this 400~by~400~$\mu$m$^2$
large-area SNSPD at 1550 nm wavelength ($\sim$0.8~eV). When the detector was
illuminated, the count rate rose at a bias  
current of 3~$\mu$A. Counts initially grew with the current and the device saturated at a bias current of 4.5~$\mu$A. At this bias current, the count rate with the laser light turned off (background count rate) was below 100 cps. The maximum background count rate was
measured at a point just below the transition to the resistive state,
at $10^3$~cps.  

The measurement of dark counts
was performed in an apparatus with several layers of shielding and the optical fiber connection removed, 
at 4.5 $\mu$A of bias current for $10^4$~s. No dark counts were observed over this period, suggesting a dark count
rate below 100 counts per microsecond. These measurements will be used
below to place bounds on DM interactions.

\section{Reach}

Our results for the reach of superconducting nanowires into the
parameter space of DM-electron scattering are shown in Fig.~\ref{fig:scat}. We
follow the analyses of Refs.~\cite{Hochberg:2015pha,Hochberg:2015fth}
for rate computation in superconducting targets, with the appropriate
modifications to Fermi energies $E_F$ and the density $\rho$ of target
materials that are typically used for the superconductors of
SNSPDs. Details of the scattering rate computation can be
found in the Appendix. 
Our results for niobium nitride~(NbN) and WSi targets use $\rho_{\rm NbN}=8.4$~g/cm$^3$ and $\rho_{\rm WSi}=9.3$~g/cm$^3$, respectively, and in both cases we use $E_F=7$~eV . 
(We note that
while the Fermi surface of WSi is not a perfect sphere, the
calculations performed here are intended to provide a proxy to guide
future experiments; we have thus assumed a spherical Fermi surface in
the case of WSi.)  

The left panel shows the reach for scattering via a
light mediator, with the commonly used reference momentum
$q_{\rm ref}=\alpha\, m_e$ defining the reference cross section
$\bar \sigma_e$, while the right panel shows the reach when scattering
via a heavy mediator (see Appendix for all definitions).  The solid
colored curves show the 95\% Confidence Level (C.L.) projected reach, corresponding to 3 signal
events, for SNSPDs with various amounts of exposures and thresholds, assuming a dynamic range
of 3 orders of magnitude. 
We also show the projected reach for a kg-year exposure of a silicon target with
single-electron sensitivity~\cite{Essig:2015cda} and of an aluminum target with a 10~meV
threshold TES~\cite{Hochberg:2015pha,Hochberg:2015fth}, along with  constraints from the Xenon10~\cite{Essig:2012yx},
SENSEI~\cite{Abramoff:2019dfb} and SuperCDMS~\cite{Agnese:2018col}
experiments.

The 95\%~C.L. bound on DM-electron scattering placed by the 
4.3~nanogram prototype WSi device with the 0.8~eV energy threshold
presented in this work, which showed no dark counts in 10~thousand
seconds of exposure, is shown by the black solid curve.
While
the bound from this prototype nanowire on DM-electron scattering is not yet competitive with
those from the other experiments, it is impressively placed using a
tiny mass-time exposure on a surface run. For comparison, the
projected reach of a WSi target with an exposure similar to that of the
SENSEI data, 0.177g-day, is also shown,
demonstrating the strength of our proposal. As is evident, larger
exposures combined with low thresholds will enable superconducting
nanowires to quickly probe uncharted parameter space of DM scattering.

\begin{figure}[th!]
\includegraphics[width=.48\textwidth]{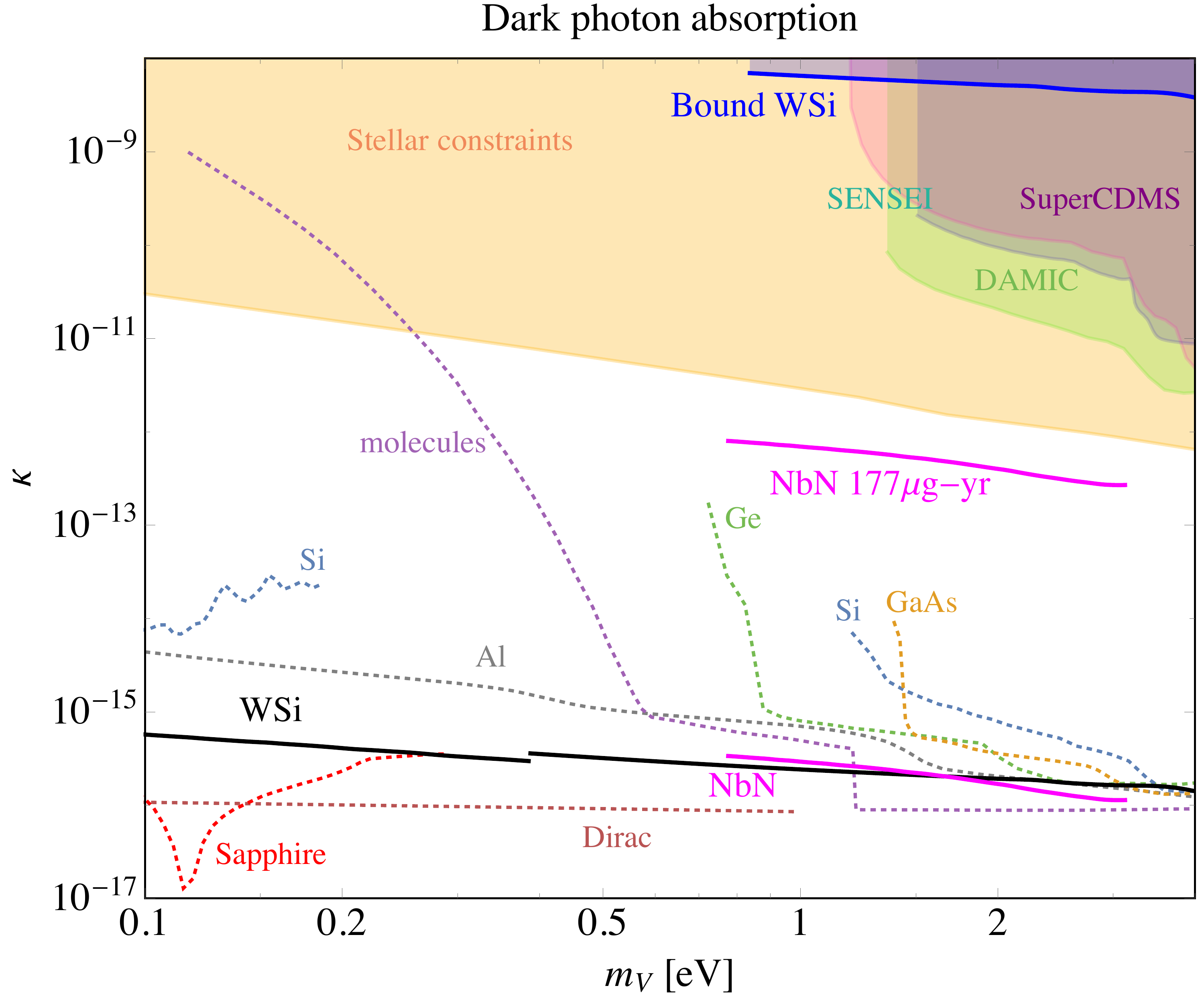} 
\caption{
\label{fig:abs}
Expected reach for absorption of relic dark photons. The solid blue
curve labeled `Bound WSi' indicates the new bound placed by our
prototype device in this work, which showed no dark counts in 4.3~ng over $10^4$~seconds. The projected reach for a NbN target SNSPD
with $177\;\mu{\rm g}$-yr (corresponding to a 10 by 10~cm$^2$ area with
4~nm thickness and a 50\% fill factor), and kg-year exposures is shown
(solid magenta curves); NbN reach into lower masses than depicted should be
possible and can be estimated from lower energy data should it become
available. Also shown is our projected reach for a kg-yr exposure of a WSi SNSPD (solid black). The reach for a kg-yr exposure of
aluminum superconductors~\cite{Hochberg:2016ajh}, semiconductors such
as germanium and silicon~\cite{Hochberg:2016sqx}, Dirac
materials~\cite{Hochberg:2017wce}, molecules~\cite{Arvanitaki:2017nhi}
and polar crystals such as GaAs and Sapphire~\cite{Griffin:2018bjn}
are given as well (dotted curves). Constraints from stellar
emission~\cite{An:2013yua,An:2014twa} are indicated (shaded
orange), along with terrestrial constraints from
SuperCDMS~\cite{Agnese:2018col} (shaded purple), 
DAMIC~\cite{Aguilar-Arevalo:2016zop} (shaded green) and SENSEI~\cite{Abramoff:2019dfb} (shaded turquoise). Unless otherwise stated,
projected reach refers to kg-yr exposure.}
\end{figure}

In addition to probing DM-electron scattering, SNSPDs can simultaneously probe absorption of relic particles that interact with electrons. As an example, we consider a relic dark photon that is kinetically mixed with the ordinary photon. Effectively, such a dark photon interacts with electrons in a similar manner to the photon, but with the interaction suppressed by the size of the kinetic mixing, $\kappa$ (see Appendix for further details).

Our results for relic dark photon absorption in SNSPDs are shown in
Fig.~\ref{fig:abs}. We use low-energy photon absorption data for
NbN~\cite{Marsili2012} and WSi~\cite{Marsili2013, Yamada2008}, and
translate it to the expected reach on the size of kinetic mixing~$\kappa$ between the photon and dark photon field strengths as a function of the dark photon mass $m_V$. Details of
the absorption rate computation can be found in the Appendix. We show the
resulting 95\% C.L. expected reach, corresponding to 3 signal events,
for a kg-year exposure of NbN and WSi target SNSPDs. For NbN, we additionally present the reach of a 177$\mu$g-yr exposure; we further note that reach into lower masses than depicted is possible
and can be estimated should lower energy data become
available. Also shown are constraints from SuperCDMS~\cite{Agnese:2018col}, 
DAMIC~\cite{Aguilar-Arevalo:2016zop} and SENSEI~\cite{Abramoff:2019dfb}, along with stellar emission
constraints~\cite{An:2013yua,An:2014twa}. The projected reach for kg-year exposures of germanium
and silicon~\cite{Hochberg:2016sqx}, superconducting
aluminum~\cite{Hochberg:2016ajh}, Dirac
materials~\cite{Hochberg:2017wce}, polar crystals such as GaAs and
shapphire~\cite{Griffin:2018bjn} as well as molecular
targets~\cite{Arvanitaki:2017nhi} are likewise indicated.

The 95\%~C.L. bound on relic dark photons placed by the data of the 4.3~ng prototype WSi
device in $10^4$~s with a 0.8~eV threshold, 
presented in this work, 
is shown by the
solid blue curve. 
 Remarkably, despite the small device size and short exposure time of our experiment, it places the strongest  terrestrial constraint to date on dark photons with sub-eV masses. 

\section{Summary}

We have proposed the use of superconducting nanowires as sensitive
targets and detectors for light dark matter, and demonstrated the
power of this approach. 
We have found that absorption of bosonic DM with masses above
the superconducting gap of ${\cal O}({\rm  meV})$ and scattering of DM in the keV to GeV mass range are both
promising and complementary to other existing proposals in these mass
ranges. An existing prototype nanowire already places meaningful
bounds on the parameter space, including the strongest terrestrial
constraints to date on dark photon absorption in the sub-eV to few-eV
mass range.

The results presented here suggest that
further work, both theoretical and experimental, is warranted to
determine the viability of using SNSPDs for this goal.  In
the Appendix, we discuss several issues raised by our
results, in particular: ({\it i})~what are the ideal device
characteristics that should be targeted; and ({\it ii})~what are the
prospects for scaling the detectors to masses large enough to
substantially extend the reach of current searches.

With low thresholds and low dark count rates, superconducting
nanowires have the potential to impact the direct detection landscape
on relatively short time scales. We hope this work serves as a
stimulant for broad cooperation between the quantum information and
fundamental physics communities, such that meaningful progress can
rapidly be made towards understanding the basic constituents of
nature.

\mysections{Acknowledgments} We thank Tom Dvir, Nadav Katz and Hadar
Steinberg for useful discussions that led to this work, and Eric David
Kramer and Eric Kuflik for many helpful discussions. We thank Phil
Mauskopf for providing the cryogenic amplifiers used in this
experiment. KKB thanks Asimina Arvanitaki and Ken van Tilburg for
helpful discussions. The work of YH is supported by the Israel Science
Foundation (grant No. 1112/17), by the Binational Science Foundation
(grant No. 2016155), by the I-CORE Program of the Planning Budgeting
Committee (grant No. 1937/12), by the German Israel Foundation (grant
No. I-2487-303.7/2017), and by the Azrieli Foundation. The work of KKB
was supported in part by the DOE under the QuantiSED program, grant
number DE-SC0019129.

\section{appendix}
%

\subsection{Prototype Device Details}

Here we provide the technical details of the prototype device used in this paper to set new bounds on dark matter scattering and
absorption. The device used in the experiment was fabricated
from 7-nm thick WSi film which was deposited on a silicon substrate at
room temperature by using an RF co-sputtering
technique~\cite{Baeka2011}. Additionally, a thin 2-nm Si layer was
$in-situ$ deposited on top of the WSi film to prevent oxidation of the
superconductor.  The critical temperature of the sputtered film was
measured at the point where $R = R_{\rm 20}$/2 ($R_{\rm 20}$ is the resistance at 20 K)  and was found to be
4.08~K. 

To pattern the nanowires, electron-beam lithography was used with
high-resolution positive e-beam resist. The ZEP 520~A resist was spin
coated onto the chip at 5000~rpm which ensured a thickness of
335~nm. The resist was exposed to a 125~keV electron beam with an area
dose density of 500~$\mu$C/cm$^{2}$. After exposure, the resist was
developed by submerging the chip in O-xylene for 80~s with
subsequent rinsing in the 2-propanol stopper. The ZEP 520~A pattern
was then transferred to the WSi by reactive ion etching in CF$_4$ at
50~W for 5 minutes.

To electrically and optically characterize the fabricated device, we
designed an experimental setup using a single shot type He-3
cryostat. The device was mounted on the sample holder using a contact
glue. The holder was placed on a 300-mK cold stage and was
shielded to reduce the effect of background radiation on the detector
noise. A low-temperature bias tee decoupled the high-frequency path
from the DC bias path. The high-frequency signal was carried out of
the cryostat by stainless-steel rigid coaxial cables, while DC bias
was provided via a pair of twisted wires connected to a
low-noise voltage source. The signal was amplified at the 4~K stage by
a cryogenic low-noise amplifier with the total gain of 56~dB and then
sent to a pulse counter. The optical single mode fiber feeds light from the
1550-nm CW laser into the cryogenic apparatus though a vacuum
feedthrough and is mounted on a stage above the sample surface. An optical attenuator was used to ensure the single-photon counting regime.

\subsection{Dark Matter Scattering Rate}\label{app:ratescat}

Here we review the scattering rate of dark matter with electrons in a superconducting
target, used to obtain our parameter space reach plot in Fig.~\ref{fig:scat}. Denoting
the 4-momentum of the DM initial and final state particle by $P_1$ and $P_3$, the
initial and final states of the electron by $P_2$ and $P_4$, and the
momentum transfer $q=(E_{\rm D}, {\bf q})$, the scattering rate is given
by~\cite{Hochberg:2015pha,Hochberg:2015fth}
\begin{eqnarray}
 \label{eq:response}
 \langle n_e\sigma v_{\rm rel}\rangle&=&\int\frac{d^3
p_3}{(2\pi)^3}\frac{ \langle|\mathcal {M}|^2\rangle}{16 E_1 E_2 E_3 E_4}\ S(E_{\rm D},|{\bf q}|)\,,\\
S(E_{\rm D},|{\bf q}|)&=&2\int\frac{d^3 p_2}{(2\pi)^3}\frac{d^3
p_4}{(2\pi)^3}(2\pi)^4\delta^4(P_1+P_2-P_3-P_4)\nonumber\\
&&\quad \quad \quad \times f_2(E_2)(1-f_4(E_4)),\nonumber
\end{eqnarray}
where $E_{\rm D}$ is the deposited energy, $ \langle|\mathcal
{M}|^2\rangle $ is the squared scattering matrix element summed and
averaged over spin, and $f_i(E_i)=[
1+\exp(\frac{E_i-\mu_i}{T})]^{-1}$ is the Fermi-Dirac distribution
of the electrons at temperature $T$, with $\mu_i$ the chemical potential (equal to the Fermi energy at zero temperature). The total rate (in number of events, per unit time per unit mass) is then 
\begin{eqnarray}
 \label{eq:RateDM}
E_{\rm D}\frac{dR_{\rm DM}}{d E_{\rm D}}=\int d v_{\rm DM} f_{\rm MB}(v_{\rm DM})
E_{\rm D}\frac{d\langle n_e\sigma v_{\rm rel}\rangle}{d
E_{\rm D}}\frac{1}{\rho}\frac{\rho_{\rm DM}}{m_{\rm DM}}\,\,\,
\end{eqnarray}
where $\rho$ is the mass density of the target material, $\rho_{\rm DM} =
0.3\ \textrm{GeV}/\textrm{cm}^3$ is the DM mass density, and $v_{\rm rel}$ is the relative velocity between the electrons and the DM. The 
DM velocity distribution $ f_{\rm MB}(v_{\rm DM})$ is taken to be a modified Maxwell Boltzmann distribution with root mean square velocity $v_0=220\ \textrm{km}/\textrm{s}$,
with a cut-off at the escape velocity $v_{\rm esc}=500\ \textrm{km}/\textrm{s}$. 

The scattering cross section of DM and electrons is related to the matrix element squared via $\sigma_{\rm scatter} =  \frac{\langle|\mathcal{M}|^2\rangle }{16\pi E_1 E_2 E_3 E_4}\mu_{e {\rm DM}}^2$, with $\mu_{e {\rm DM}}$ the reduced DM-electron mass, and we define a reference cross section $\overline \sigma_e$ as is common in the literature with fixed momentum transfer $q_{\rm ref} = \alpha  m_e$. Namely, 
\beq
\overline\sigma_e \equiv \frac{\mu_{e {\rm DM}}^2}{16\pi m_{\rm DM}^2 m_e^2}\langle|\mathcal{M}(q)|^2\rangle |_{q^2=\alpha^2 m_e^2}\,,\\
\langle|\mathcal{M}(q)|^2\rangle = \langle|\mathcal{M}(q)|^2\rangle |_{q^2=\alpha^2 m_e^2} \times |F_{\rm DM}(q)|^2\,,
\eeq
with $F_{\rm DM}(q)$ the form factor that encompasses the momentum
dependence of the cross section. $\overline \sigma_e$ is then equal to
the non-relativistic DM-electron elastic cross section with the
reference momentum transfer $q_{\rm ref}=\alpha m_e$. For scattering
processes that occur via the exchange of a heavy mediator between the
DM and the electrons, $F_{\rm DM}=1$, while for a light mediator
exchange, $F_{\rm DM}=q_{\rm ref}^2/q^2$.

Our results for the reach of superconducting nanowires into the
reference cross section $\overline \sigma_e$ of DM-electron scattering parameter space, for
heavy and light mediators, are shown in Fig.~\ref{fig:scat}.

\subsection{Dark Photon Absorption Rate}\label{app:rateabs}

Here we review how we obtain the absorption rate of a relic dark
photon in a given material, which leads to our
Fig.~\ref{fig:abs}, via the optical response of the material. We
consider the Lagrangian of a dark photon with an induced kinetic mixing with the photon
(sourced by a kinetic mixing with the hypercharge gauge boson),
\beq\label{eq:LHidden} {\cal L}\supset -\frac{\kappa}{2}
F_{\mu\nu}V^{\mu\nu}\,, \eeq
where $F^{\mu\nu}$ and $V^{\mu\nu}$ are the field strengths for the
photon~$A$ and dark photon~$V$, respectively: $F_{\mu\nu}=\partial_\mu A_\nu-\partial_\nu A_\mu$, and $V^{\mu\nu}$ is similarly defined with the replacement $A^\mu \to V^\mu$ (here $\mu, \nu=0,...4$ are Lorentz indices). The effective mixing angle in a
medium is given by
\beq\label{eq:kappaeff}
\kappa_{\rm eff}^2 =\frac{\kappa^2 m_V^4}{\left[m_V^2-{\rm Re}\,\Pi(m_V)\right]^2+\left[{\rm Im}\, \Pi(m_V)\right]^2}\,,
\eeq
where $m_V$ is the mass of the dark photon, and 
 $\Pi$ is the polarization tensor of the material,
related to the optical conductivity of the material $\hat \sigma$ through
\beq\label{eq:pisig}
\Pi(\omega)\approx - i \hat \sigma \omega\,,
\eeq
with the conductivity related to the complex index of refraction
$\hat n$ by $\hat n^2 = 1 + i \hat \sigma/\omega$.  We relate the
absorption of dark photons to that of ordinary photons to obtain the
absorption rate for dark photons in the given material. For absorption
processes, the deposited energy is $\omega\approx m_V$.  The result is that the absorption
rate in counts per unit mass per unit time is %
\beq\label{eq:rateHidden} R=\frac{1}{\rho}\frac{\rho_{\rm DM}}{m_{\rm
    V}}\kappa_{\rm eff}^2 {\rm Re}\left[\hat \sigma(m_V)\right]\,, \eeq
where $\kappa_{\rm eff}^2$ is given by Eq.~\eqref{eq:kappaeff},  $\rho$ is the mass density of the target and
$\rho_{\rm DM}=0.3\;{\rm GeV}/{\rm cm}^3$ is the DM mass density, as
before.

Our results for the reach of superconducting nanowires into the
parameter space of dark photon absorption are shown in
Fig.~\ref{fig:abs}.

\subsection{Detector Considerations}

We now consider issues associated with scaling the
detectors to larger areas, higher efficiencies, lower dark count
rates and lower energies.  All of these elements correspond directly
to improved system reach. Here, we propose a metric that can be used to
quickly compare detector technologies, and discuss the prospect of
scaling that metric by increasing the detector area.

\mysection{Dark Counts Figure of Merit} A useful figure of merit
(FOM) for DM detection with these detectors is the product of the
device area $A$ with the detection efficiency $\eta$ divided by the
dark count rate (DCR), $\mathrm{FOM} = \eta A / \mathrm{DCR}$. This
FOM is dependent on wavelength, bias current, and film thickness, but is still relevant when comparing within existing detector families where these parameters are held relatively constant. Recent
work~\cite{Wollman2017} has demonstrated a detector with FOM of
$2.7\times 10^{-5}\,\mathrm{m}^2/\mathrm{cps}$ for \SI{4}{eV} light
measured at $T = 800$~mK.

The main parameters available to optimize this FOM are the bias
current, the film thickness, and the detector
geometry/design. Typically, higher bias current will increase the
efficiency, but would also potentially increase the DCR, thus an intermediate value
is desirable.  In addition, thicker films will result in increased
mass, but at the expense of reduced efficiency.  Additionally, lower-energy photons are likely to be detectable with reduced efficiency in a thicker film.
These complicated trade-offs mean that the detector design space used
here is unlikely to be optimal, and further improvements are possible.

\mysection{Scaling} Any reasonable effort at DM detection based on the
proposed technique will require larger areas of detector to be
manufactured. Recent work from Ref.~\cite{Korneeva2018} 
has suggested that nanometer-length-scale device widths may not be
required in order to realize SNSPDs (more properly then, SSPDs, as the
wires are no longer true `nanowires').  In that case, optical
lithography could be used exclusively in their fabrication and as such
mass production at the wafer scale of large-area devices would be
quite viable.  An academic laboratory could readily produce in the
course of a year a thousand 200-mm-diameter wafers, resulting in a
total detector mass of 1.3~g.  An industrial effort could realize many
times that number.  At that point, the main challenge would be testing
and packaging, rather than simply manufacture. With proper design,
planning and resources, target masses in the kg-range may be feasible.

\bibliography{bibliocoolnew}{}

\end{document}